\documentclass[letter,traditabstract,longauth]{aa}
\usepackage{graphicx}
\usepackage{txfonts}

\newcommand\simgt{\lower.5ex\hbox{$\;\buildrel>\over\sim\;$}}
\newcommand\simlt{\lower.5ex\hbox{$\;\buildrel<\over\sim\;$}}

 \newcommand{\mic}{$\mu$m}
 \newcommand{\mics}{$\mu$m~}

\def\cone {\ifmmode{{\rm C}{\rm \small I}(^3\!P_1\!-^3\!P_0)}
     \else{C\ts {\scriptsize I}{\small$(^3\!P_1\!-^3\!\!\!P_0)$}}\fi}
\def\ctwo {\ifmmode{{\rm C}{\rm \small I}(^3\!P_2\!-^3\!P_1)}
     \else{C\ts {\scriptsize I}{\small$(^3\!P_2\!-^3\!\!\!P_1)$}}\fi}

\def\tex {\ifmmode{{T}_{\rm ex}}\else{$T_{\rm ex}$}\fi}
\def\tmb {\ifmmode{{T}_{\rm mb}}\else{$T_{\rm mb}$}\fi}
\def\ci     {\ifmmode{{\rm C}{\rm \small I}}\else{C\ts {\scriptsize I}}\fi}
\def\hi     {\ifmmode{{\rm H}{\rm \small I}}\else{H\ts {\scriptsize I}}\fi}
\def\hh     {\ifmmode{{\rm H}_2}\else{H$_2$}\fi}

\def\ts     {\thinspace}
\def\kms    {\ifmmode{{\rm \ts km\ts s}^{-1}}\else{\ts km\ts s$^{-1}$}\fi}
\def\msol   {\ifmmode{{\rm M}_{\odot}}\else{M$_{\odot}$}\fi}
\def\lsol   {\ifmmode{{\rm L}_{\odot}}\else{L$_{\odot}$}\fi}
\def\zsol   {\ifmmode{{\rm Z}_{\odot}}\else{Z$_{\odot}$}\fi}
\def\etal   {{\rm et\ts al.\ts}}

\def\watwo  {{\rm H$_2$O$_p$}(2,0,2-1,1,1)}
\def\hls  {{\rm HLSJ091828.6+514223}}

\begin{document}

\title{A bright z=5.2 lensed submillimeter galaxy in the field of Abell 773}

\subtitle{HLSJ091828.6+514223}

\author{F. Combes        \inst{1}
      \and       M. Rex        \inst{2}
      \and       T. D. Rawle        \inst{2}
      \and       E. Egami        \inst{2}
      \and       F. Boone        \inst{3}
      \and       I. Smail       \inst{4}
      \and       J. Richard       \inst{5}
      \and       R.J. Ivison       \inst{6}
      \and       M. Gurwell       \inst{7}
      \and       C.M. Casey       \inst{8}
      \and       A. Omont        \inst{9}
      \and       A. Berciano Alba      \inst{10}
      \and       M. Dessauges-Zavadsky        \inst{11}
      \and       A.C. Edge       \inst{4}
      \and       G.G. Fazio       \inst{7}
      \and       J-P. Kneib       \inst{12}
      \and       N. Okabe       \inst{13}
      \and       R. Pell\'o       \inst{3}
      \and       P. G. P\'erez-Gonz\'alez       \inst{14}
      \and       D. Schaerer       \inst{11,3}
      \and       G.P. Smith       \inst{15}
      \and       A.M. Swinbank       \inst{4}
      \and       P. van der Werf      \inst{16}
}

\offprints{F. Combes}

\institute{Observatoire de Paris, LERMA, CNRS, 61 Av. de l'Observatoire, 75014 Paris, France
                \email{francoise.combes@obspm.fr}
\and Steward Observatory, University of Arizona, 933 North Cherry Avenue, Tucson, AZ, 85721, USA
\and  Universit\'e de Toulouse, UPS-OMP, CNRS, IRAP, 9 Av. colonel Roche, BP 44346, 31028, Toulouse Cedex 4, France
\and Institute for Computational Cosmology, Durham University, South Road, Durham, DH1 3LE, UK
\and CRAL, Universit\'e Lyon-1, 9 Av. Charles Andr\'e, 69561 St Genis Laval, France
\and UK Astronomy Technology Centre, Royal Observatory, Blackford Hill, Edinburgh, EH9 3HJ, UK
\and Harvard-Smithsonian Center for Astrophysics, 60 Garden Street, Cambridge, MA 02138, USA 
\and Institute for Astronomy, University of Hawaii, 2680 Woodlawn Dr, Honolulu, HI 96822, USA
\and Institut d'Astrophysique de Paris, UPMC and CNRS, 98 bis Bd. Arago, 75014 Paris, France
\and ASTRON, P.O. Box 2, NL-2990 AA Dwingeloo, The Netherlands
\and Geneva Observatory, Universit\'e de Gen\`eve, 51 chemin des Maillettes, 1290, Versoix, Switzerland
\and LAM, CNRS- Universit\'e Aix-Marseille, 38 rue F. Joliot-Curie, 13388, Marseille Cedex 13, France
\and Academia Sinica Institute of Astronomy and Astrophysics (ASIAA), P.O. Box 23-141, Taipei 10617, Taiwan
\and Dep. de Astrofisica, Facultad de CC. Fisicas, Universidad Complutense de Madrid, 28040 Madrid, Spain
\and School of Physics and Astronomy, University of Birmingham, Edgbaston, Birmingham B15 2TT, UK
\and Leiden Observatory, Leiden University, PO Box 9513, 2300 RA, Leiden, The Netherlands
             }

\date{Received December 28, 2011 / Accepted January 16, 2012}

\abstract{During our {\it Herschel} Lensing Survey (HLS) of massive galaxy clusters,
we have discovered an exceptionally bright source behind the z=0.22 cluster Abell 773,
which appears to be a strongly lensed submillimeter galaxy (SMG) at z=5.2429. This source is unusual
compared to most other lensed sources discovered by {\it Herschel} so far, because of
its higher submm flux ($\sim$200mJy at 500\mic) and its high redshift. 
The dominant lens is a foreground z=0.63 galaxy, not the cluster itself. The source has a 
far-infrared (FIR) luminosity of  L$_{\rm FIR}$ = 1.1 10$^{14}/\mu$ \lsol,
where $\mu$ is the magnification factor, likely $\sim$ 11.  
We report here the redshift identification through CO lines with
the IRAM-30m, and the analysis of the gas excitation, based on 
CO(7-6), CO(6-5), CO(5-4) detected at IRAM and the CO(2-1) at the EVLA. 
All lines decompose into a wide and
strong red component, and a narrower and weaker blue component,
540\kms apart.  Assuming the ultraluminous galaxy (ULIRG) CO-to-H$_2$ conversion
ratio, the \hh\,  mass is  5.8 10$^{11}/\mu$ M$_\odot$, of which one third is 
in a cool component. From the \ctwo\, line we derive a
 \ci/\hh\, number abundance of 6 10$^{-5}$ similar to that in other
ULIRGs. The \watwo \ts line is strong only in the red velocity component,
with an intensity ratio $ I(H_2O)/I(CO) \sim$ 0.5, suggesting a strong local
FIR radiation field, possibly from an active nucleus (AGN) component. 
We detect the [NII]205\mics line for the first time at high-z.
It shows comparable blue and red components, with a strikingly 
broad blue one, suggesting strong ionized gas flows.

}

\maketitle

\section{Introduction} 

At high redshift, z$\sim$1--3, dusty luminous infrared galaxies 
are thought to dominate the history of cosmic star formation
(Dole \etal 2006; Wardlow \etal 2011).
At these epochs, star formation was occuring mainly
in massive galaxies, a tendency dubbed downsizing in galaxy formation scenarios
(e.g. Cowie \etal 1996; Heavens \etal 2004;
P\'erez-Gonzalez \etal 2008; Magnelli \etal 2009, 2011).
These massive starbursts might correspond to the formation phase of the luminous
ellipticals seen in high-density regions today (Genzel \etal 2003;
Swinbank \etal 2006). They are therefore
essential for understanding the main processes of galaxy formation and evolution.
In particular, atomic and molecular lines are crucial clues in these systems for
tackling star formation and AGN (Active Galaxy Nucleus) activities (Riechers \etal 2010, 2011; Danielson \etal 2011). 

Owing to the negative K-correction, the number of submillimeter  galaxies (SMGs) does not need to 
peak at z$\sim$1--3, indeed, if they  exist, they should be as easy to detect at z=8 as they are at  z=1--3.
Their observed dearth 
is probably due to evolution of gas fraction and halo mass  (Lacey \etal  2010; Hickox \etal 2011).
 To be able to make a census of z$\sim$ 5 SMG and  compare their number with lower-z SMGs,
we use {\it Herschel} to identify candidate high-z SMGs and strong lensing 
 to provide the signal-to-noise and a unique spatial resolution. 
This technique has now proven quite successful in the study
of high-z SMG (e.g. Smail et al 1997;
Pell\'o \etal 1999; Swinbank \etal 2010; Lestrade \etal 2010;
Negrello \etal 2010; Lupu \etal 2011; Cox \etal 2011).

Strongly lensed sources are quite rare on the sky, and are serendipitously discovered
in large surveys, such as the South Pole Telescope survey (Vieira \etal 2010), 
the {\it Herschel} HERMES (Oliver \etal 2010), or the huge area {\it Herschel}-ATLAS 
(H-ATLAS) survey (Eales \etal 2010).
We have undertaken a more targeted survey with {\it Herschel},
focusing on galaxy clusters. Our key project,
"{\it Herschel} Lensing Survey" (HLS-deep), has the goal to deeply image  44 massive clusters 
of galaxies with PACS and SPIRE (Egami \etal 2010; Rex \etal 2010), and continues
with a SPIRE snapshot survey (HLS-snapshot) of $\sim$ 300 clusters.
The unique aim of the HLS survey is to find
cluster-lensed sources, as opposed to galaxy-lensed objects found in field surveys 
such as H-ATLAS.  These cluster-lensed sources suffer less from differential amplification
than galaxy-lensed sources, while the dense environment of clusters 
increases the probability of galaxy lensing (Smail et al 2007).

The HLS-deep survey is now fully executed and 
the source behind Abell 773 is the strongest source at 500\mic, among the 44 clusters surveyed.
Together with the $z=2.78$ lensed galaxy in the Bullet cluster (Rex et al. 2010), 
it is one of only two HLS-deep SPIRE sources with a peak flux density 
at 500\mics above 100 mJy.
It is located far from the cluster center ($\sim$4.5' = 1 Mpc), and likely lensed by a single galaxy.
 
We present in this letter the continuum and molecular line observations
of the Abell 773 SMG, deriving its gas excitation and physical properties.
\S 2 reports the discovery of the SMG from our SPIRE and SMA data,
IRAM spectroscopic observations to derive its redshift, with
detections of CO, \ci, H$_2$O, [NII] lines
and CO(2-1) with EVLA.
In \S 3 we model the data to infer the dust and gas content and its general properties.
For distances we adopted the $\Lambda$-CDM concordance
model:  H$_0$ = 71 km/s/Mpc,
$\Omega_{\small M} $ = 0.27 and $\Omega_{\Lambda}$ = 0.73.

\section{Observations} 
\label{obser}

\subsection{Herschel and SMA} 
\label{spire-sma}

SPIRE observations were conducted using the "large-map mode", with orthogonal 
cross-linked scans producing 17$'$x17$'$ coverage in each of the three SPIRE bands.  
The maps include 20 repetitions with a total observing time of 1.7 hours, giving an on-source 
integration time of $\sim$2000 s.  The data were reduced using the Herschel interactive pipeline 
environment (HIPE v5.0).  The maps were produced with the standard naive map maker, 
but include additional turnaround data with scan speeds $>0.5$\arcsec/s to increase 
the depth of the outer regions of the maps.  For more details see Egami \etal (2010).

\begin{figure}[!t]
\resizebox{8cm}{!}{\includegraphics[angle=-90]{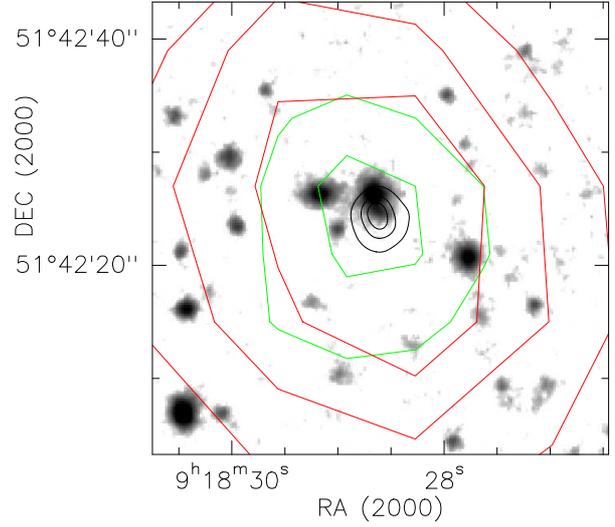}}
\caption{SPIRE 250\mics  (green, beam 18\arcsec), 500\mics (red, 37\arcsec) and SMA 880\mics 
(black, 2\arcsec) overplotted on the Subaru (suprime cam) 
V-image of \hls. The latter shows the lensing galaxy at the center, slightly north of the SMA peak.
The SMA contours start at 1$\sigma$ and increase in steps of 4$\sigma$.
The  EVLA CO(2-1) contours are peaked  at the same position and have the same N-S elongation.
}
\label{fig:spire-sma}
\end{figure}

\hls\, was discovered in the SPIRE map of the z=0.22 cluster Abell 773 (Barrena \etal 2007), 
shown in Figure \ref{fig:spire-sma}. Its position is
$\alpha_{2000}=09^h 18^m 28.6^s$ and $\delta_{2000}=51^{\circ} 42' 23.3$\arcsec,
too far away to be covered by PACS or {\it Spitzer} images of the cluster.
It lies close to an optical source (Fig \ref{fig:spire-sma}), which is likely the lensing galaxy. 
\hls\, has an SED (spectral energy distribution) peaking at 500\mic, with F$_{500}$=203 mJy,
suggesting a high redshift.
The mm/submm contribution of the foreground galaxies is negligible.

The Submillimeter Array (SMA) observed the source at 235 GHz
  (1.3 mm) on 15 Jun 2011 with precipitable water vapor (pwv)
between 1.5 and 4mm during 62min on-source,
with 8 GHz total bandwidth.
  The source was reobserved at 341 GHz (880\mic) on
  9 Dec 2011, during 330 minutes on-source and pwv of 1.5mm.
  Phase and amplitude gain variations
  with time were monitored using quasars 0927+390 and 0920+446, and
  the flux scale was determined using Titan (Table \ref{tab:continuum}).
  The source is slightly resolved at 1.3mm by the 4.6x2.8\arcsec\ts beam, and 
resolved at 880\mic by the 2.1x2.0\arcsec\ts beam, with
  an N/S extension (Fig \ref{fig:spire-sma}).

\subsection{IRAM and EVLA} 
\label{iram}

\begin{figure}[!t]
\resizebox{8cm}{!}{\includegraphics[angle=0]{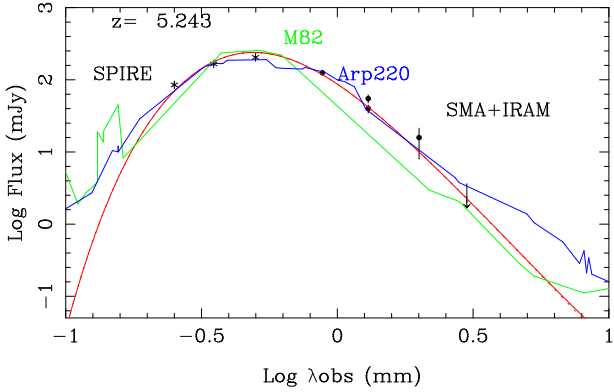}}
\caption{SPIRE (250, 350 and 500\mic), SMA (0.88 and 1.3mm)  and IRAM (1.3, 2 and 3mm) photometric data for 
\hls, superposed on our model (red line) of a single big
 grain component at T$_d$=52K ($\beta=2$). The M82 and Arp220 SED from NED are also plotted.
Our source has an SED very similar to Arp 220.
}
\label{fig:SED}
\end{figure}

To measure the redshift of the source, we used the strategy 
developed by Weiss \etal (2009)
and Lestrade \etal (2010) with the multi-band heterodyne receiver EMIR
at the IRAM 30-m telescope. At the time of observations (26-27 Sep and 6 Oct 2011),
one setup with the 3mm receivers provided 7.43~GHz of instantaneous, dual linear
polarization  bandwidth. We then covered the 35 GHz, from 80 to 115 GHz, in five setups with 
some overlaps,  integrating about one hour per setup, in wobbler-switching mode.
The precipitable water vapor was between 2 and 5mm,
and the system temperatures was 90-100K at 3mm and 200K at 1.3mm. We used two backends,
the 4MHz filter-banks, and the WILMA autocorrelator, which provides a spectral resolution of 2 MHz.
Pointing and focus offsets, determined once every two hours,
were more accurate than 3\arcsec. 
Continuum levels were obtained at 2mm and 1.3mm (and an upper limit at 3mm, cf
Fig. \ref{fig:SED} and Table  \ref{tab:continuum}). At 1.3mm, we detected
40$\pm$5mJy, slightly lower than, but still compatible with the SMA value.

 The search revealed a line only at the fourth setup, at 92.3 GHz. The fifth setup gave
a second line at 110 GHz, suggesting a redshift, which was then confirmed by tuning the 
 2mm receiver at 129 GHz, the three detected lines thus being $^{12}$CO J=5-4, J=6-5
and J=7-6 transitions, indicating a redshift of 5.2429.
Other lines were searched for, such as dense tracers HCN and HCO$^+$ at 3mm,
water at 2mm, and the [NII]205\mics  line at 1.3mm. 

Around 1\,hr of integration was also obtained in CO(2--1) using the
Expanded Very Large Array (Perley \etal 2011), with observations and
data reduction following those reported by Ivison \etal (2011). 

%__________________________________________________ One column table
   \begin{table}[!b]
      \caption[]{Flux  $S_\nu$(mJy) for \hls, with IRAM-30m (I), SMA (S) and {\it Herschel} (H)}
         \label{tab:continuum}
            \begin{tabular}{lllllll}
            \hline
            \noalign{\smallskip}
           3 mm & 2mm & 1.3mm & 0.88mm &500\mic & 350\mic & 250\mic \\
           I & I & S & S & H & H & H \\
         $<$2 & 15$\pm$7 &  55$\pm$7 & 125$\pm$8&203$\pm$9 & 168$\pm$8  &85$\pm$8  \\
            \noalign{\smallskip}
            \hline
           \end{tabular}
\begin{list}{}{}
\item[] Not including absolute flux calibration uncertainties
\end{list}
\end{table}
%---------------------------------------------

Figure \ref{fig:spectra} displays all detected lines, and results
of the Gaussian fits are reported in Table \ref{tab:lines}.

%__________________________________________________ Two column table
   \begin{table*}
      \caption[]{Observed line parameters toward \hls\,}
         \label{tab:lines}
            \begin{tabular}{l c c c c c c c c}
            \hline
            \noalign{\smallskip}
            Line & $\nu_{\rm obs}$ & \tmb & $S_\nu$ & $\Delta
      V_{\rm FWHM}$ & $I$& $V^{*}$ & $L'$/10$^{10}$   & Telescope           \\
                 & [GHz] & [mK] & [mJy] & [\kms] & [Jy \kms]&[\kms] & [K  \kms\,pc$^2$] & \\
            \noalign{\smallskip}
            \hline
            \noalign{\smallskip}
 CO(2--1) blue & 36.9280 &      & 3.1 $\pm$ 0.7   & 230 $\pm$ 60 & 0.8 $\pm$ 0.1 &  $-$530 $\pm$ 20 &17. $\pm$ 2. & EVLA \\
 red    &     &                &  4.0 $\pm$ 0.7  & 510 $\pm$ 30 & 2.2 $\pm$ 0.1 &  $-$20 $\pm$ 20 &55. $\pm$ 3. &   \\
 CO(5--4) blue & 92.3077 & 1.7 $\pm$ 0.3 & 8.6 $\pm$ 1.5  & 110 $\pm$ 30 & 1.0 $\pm$ 0.2 &  $-$540 $\pm$ 12 &4. $\pm$ 1.  & IRAM-30m\\
 red    &     &  2.7  $\pm$ 0.3 &  13.4 $\pm$ 1.5     & 540 $\pm$ 40   & 7.7 $\pm$ 0.5 &  $-$5 $\pm$ 17 &30. $\pm$ 2.&  \\
CO(6--5) blue & 110.7615 & 1.7 $\pm$ 0.4&  11. $\pm$ 2 & 160 $\pm$ 60 & 1.9 $\pm$ 0.4 &  $-$540 $\pm$ 25&5. $\pm$ 1.& IRAM-30m \\
   red  &      &  3.1 $\pm$ 0.4&  16 $\pm$ 2. & 510 $\pm$ 50 & 8.3 $\pm$ 0.6 &  0 $\pm$ 17&23. $\pm$ 2.&  \\
 CO(7-6) blue &129.2111& 2.0  $\pm$ 0.3&  9.7 $\pm$ 1.5   & 150 $\pm$ 20  & 1.5 $\pm$ 0.2 &  $-$510 $\pm$ 30 &3. $\pm$ .5 & IRAM-30m\\
  red &      & 2.9  $\pm$ 0.3&  14.3 $\pm$ 1.5   & 560 $\pm$ 20  & 8.5 $\pm$ 0.3 &  4 $\pm$ 10 &17. $\pm$ 1.& and PdBI \\
 \ctwo \ts blue & 129.6422 & 1.6 $\pm$ 0.3 & 8.3 $\pm$ 1.5 & 150 $\pm$ 20  & 1.4 $\pm$ 0.2 & $-$530 $\pm$ 9 &2.8 $\pm$ .5& IRAM-30m\\
 red &       &  1.7  $\pm$ 0.3&  8.8 $\pm$ 1.5   & 530 $\pm$ 40  & 4.9 $\pm$ 0.3 &  20 $\pm$ 14 &10. $\pm$ 1.& and PdBI \\
 \watwo  &  158.2481  & 2.2 $\pm$ 0.5 &  11 $\pm$ 2  & 400 $\pm$ 60 & 4.8 $\pm$ 0.7  & $-$17 $\pm$ 30 &6.5 $\pm$ 0.8& IRAM-30m\\
 NII blue & 234.0469&  1.6 $\pm$ 0.2&  8 $\pm$ 1 & 410 $\pm$ 90 & 3.5 $\pm$ 0.6 &  $-$620 $\pm$ 35&2.2 $\pm$ 0.4& IRAM-30m \\
 red  &   &  1.7 $\pm$ 0.2&  8.7 $\pm$ 1 & 340 $\pm$ 60 & 3.2 $\pm$ 0.5 &  $-$33 $\pm$ 30&2.0 $\pm$ 0.4 & \\
            \noalign{\smallskip}
            \hline
           \end{tabular}
\begin{list}{}{}
\item[] Data and errors are derived from Gaussian fits for the two V-components.
\item[$^{*}$] The velocity is given relative to z=5.2429 (centered on the red component of the CO(6-5) line)
\end{list}
\end{table*}

\section{Results}

\subsection{Lens modeling}
 The $V$ and $i$ images obtained with the Suprime Cam at Subaru shows
the presence of a foreground galaxy at $\sim$ 1.2\arcsec\ts north of the SMG source.
 A 3x10min spectrum was taken on KeckII with DEIMOS 600l/mm grating at 7200\AA\, with the GG455 filter, 
centered on the foreground galaxy using a 1\arcsec\ts slit at PA=14$^\circ$ in $\sim$1\arcsec\ts seeing.
The spectroscopic redshift of the lens was measured as z=0.63$\pm$0.005, 
from CaII-HK lines and the G band.
Adopting an apparent size for the lensed source of $\sim$ 2\arcsec\ts from the SMA maps, we have  
used {\sc lenstool} (Kneib \etal 1993) to produce
Einstein rings or crosses of this size, and estimate a magnification $\mu\sim$11 (sum of the images) 
for a lens at z=0.63. Since the images are not identified yet, we estimate
that 11 is the maximum amplification, whose range is 5-11.
The cluster does not boost
the magnification by more than 10\% (Richard \etal 2010).

\subsection{Dust emission}
Table \ref{tab:continuum} displays the available photometric data from SPIRE, SMA
and the IRAM-30m.
We fitted the data in Fig.\,\ref{fig:SED} with a single optically thin
big grain component at an average temperature $T_d$=52K,
an emissivity slope $\beta$=2, and a mass absorption coefficient 
$\kappa$=50 cm$^2$/g at 100\mic. 
We derived the total FIR(8-1000\mic) luminosity L$_{\rm FIR}$ = 1.1 10$^{14}/\mu$ \lsol,
and the star-formation rate SFR = 1.8 10$^{4}/\mu$ \msol/yr, applying the relation from Kennicutt (1998).  
Fitting with other dust models, such as Chary \& Elbaz (2001), leads to the 
same luminosity, within the uncertainties.
The mass of the dust is then M$_{dust}$ = 5.0 10$^9/\mu$ M$_\odot$, which for a
gas-to-dust mass ratio of 150 leads to M$_{gas}$ = 7.6 10$^{11}/\mu$ M$_\odot$.
Other fits are possible, although less good,
with lower values of $\beta$ (down to 1.5), higher dust temperatures
and lower dust masses, down by a factor 4.

\begin{figure}[!b]
\resizebox{8.4cm}{!}{\includegraphics[angle=0]{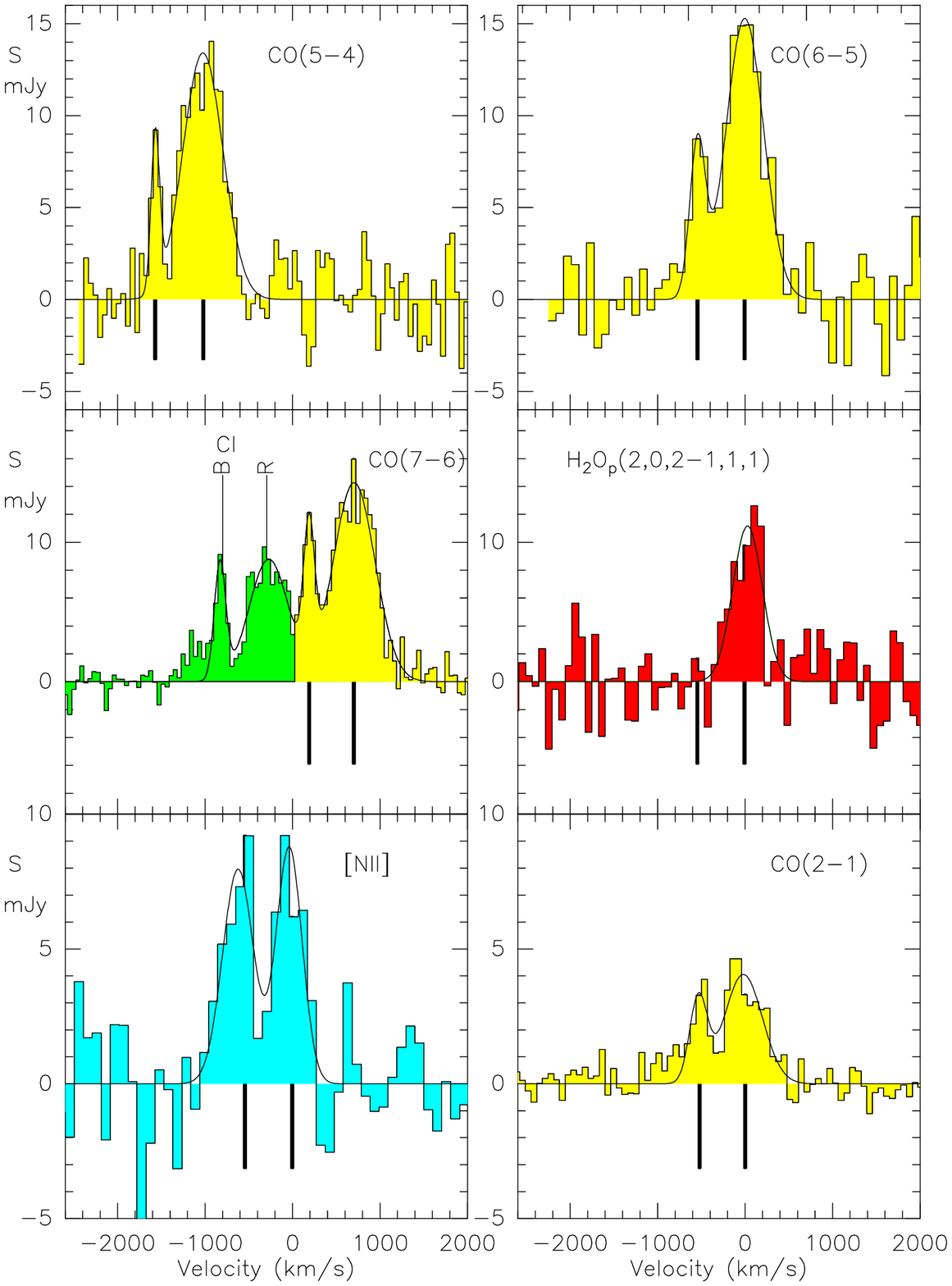}}
\caption{Four CO lines, with the \ctwo\,, 
 H$_2$O$_p$(2,0,2-1,1,1) and [NII] lines. All spectra are from
the IRAM-30m, except for the CI/CO(7-6) from the PdBI, and CO(2-1) from EVLA.
 Upper vertical lines indicate the red (R) or blue (B) components for CI,
 and lower vertical lines indicate the same for all others. All panels have
the same velocity scale (translated for CO(5-4) and CO/CI).
The blue component of the [NII] line is much stronger
and broader than for the CO lines, suggesting a strong ionized gas flow.
Exact velocities are detailed in Table \ref{tab:lines}.
}
\label{fig:spectra}
\end{figure}

\subsection{CO lines}
\begin{figure}[!t]
\centering
\resizebox{6.9cm}{!}{\includegraphics[angle=-90]{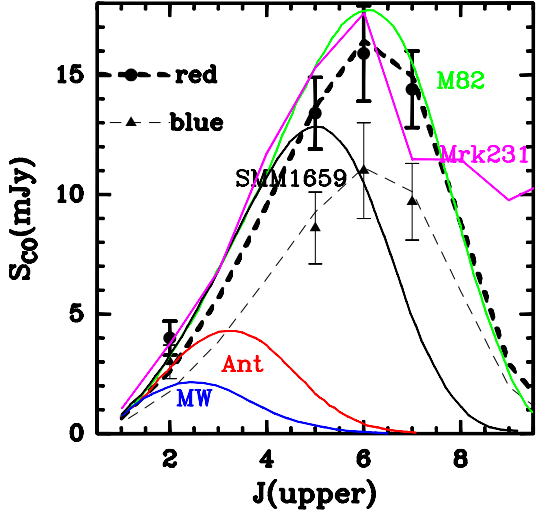}}
\caption{Peak CO line fluxes for the red 
(dots) and blue component (triangles), with the best LVG model (bold and light
 dashed lines, respectively)
computed with a density n$_{\hh}$ = 3.5 10$^3$ cm$^{-3}$, T$_{k}$ = 45K,
and a column density N(CO) = 10$^{18}$ cm$^{-2}$/\kms. The light dashed line is
related to the bold line in a ratio of 0.8 to 1.
The color lines represent CO data from other galaxies (e.g. Weiss \etal 2007; Danielson \etal 2011).}
\label{fig:SLED}
\end{figure}

Four CO lines are clearly detected in two components,
as shown in Figure \ref{fig:spectra}. The peak flux of the stronger red
component is modeled in Figure \ref{fig:SLED} with an LVG (large velocity gradient) code,
varying  the \hh\, volume density, the CO column density
and the kinetic temperature (Combes \etal 1999). The blue component has
a peak intensity lower by 25\%, whatever the $J$ level, within the
error bars. 
The three high-J CO lines are well fitted by a single excitation component,
peaking at the $J=6$ level.  We found
the best fit  with a density n$_{\hh}$ = 3.5 10$^3$ cm$^{-3}$, T$_{k}$ = 45K,
and a column density N(CO) = 10$^{18}$ cm$^{-2}$/\kms.
The solution is somewhat degenerate, between low \hh\,
density and higher CO column density, and we favor optically thick
CO emission. 
PDR (photodissociation region) models fit as well, the CO being emitted by small and dense 
optically thick clouds. Observations of higher J CO lines are required
to distinguish between PDR and XDR (Xray-dissociated region), where CO is emitted by a more extended
and diffuse hot medium (Meijerink \etal 2007).  
The observed CO(2-1) flux is 50\% higher
than predicted by the high-excitation component, suggesting the presence of
a cooler component, containing one third of the mass. 
 Adopting the conversion factor M(\hh)/L$'_{\rm CO(1-0)}$=0.8 M$_\odot$/(K \kms pc$^2$) 
of ULIRGs (ultraluminous galaxies, e.g. Solomon \etal 1997), we derive from the CO(2-1) integrated flux
an \hh\,  mass of 5.8 10$^{11}/\mu$ M$_\odot$. 
This mass is 0.8 times that derived from the dust emission, but
there are still uncertainties in the conversion ratio to apply.
 The derived ratio L$_{\rm FIR}$/L$'_{\rm CO(3-2)}\sim$ 260, compatible with 
other (U)LIRGs (Iono \etal 2009).

\subsection{Atomic carbon}
Since the \cone\, line is not available in the receiver range,
we have observed the \ctwo\, with the 30m, and confirmed it later
with the Plateau de Bure (PdBI), with a better signal-to-noise ratio (Boone \etal in prep). The latter is shown
in Fig \ref{fig:spectra}, with the two velocity components detailed in 
 Table \ref{tab:lines}.
From an adopted excitation temperature of $T_{ex}$ = 40K, and 
a total luminosity $L'_{\ctwo}$= 12.8 10$^{10}$ K  \kms\,pc$^2$,
we derive a carbon mass of M$_{\ci}$=2 10$^8/\mu$ M$_\odot$
(Weiss \etal 2005). This yields 
 a \ci/\hh\, number abundance of 6$\pm$2 10$^{-5}$, comparable
to other high-z SMGs (e.g. Danielson \etal 2011).

\subsection{Water}

ISO and {\it Herschel} observations have revealed that far-infrared H$_2$O lines
are surprisingly strong with respect to CO in ULIRGs, 
in particular those containing an AGN (Arp 220, Cloverleaf, Mrk 231,
Gonzalez-Alfonso \etal 2004, 2008; Bradford \etal 2009; van der Werf \etal 2010).
Water is thought to evaporate from grains in shocked regions,
in dense hot cores due to cosmic rays or possibly X-rays near AGN
(Gonzalez-Alfonso \etal 2010). In Mrk231, van der Werf \etal (2010)
have shown that starbursts and PDR alone cannot explain 
the high-J excitation of the CO lines, nor the H$_2$O/CO ratios, 
and a very strong FIR radiation field or
 XDR conditions related to AGN are required.
 The associations of starbursts with AGN might be even more frequent
at high redshift, and a strong H$_2$O line has been detected
at z=2.3 (Omont \etal 2011) and z=3.91 (e.g. van der Werf \etal 2011).

The water line detected in \hls\, is the highest flux line expected
from the mixed model of Gonzalez-Alfonso \& Cernicharo (1999),
including collisional excitation and far-infrared pumping. 
 We did not detect the  H$_2$O$_o$(1,1,0-1,0,1), with an upper
limit 3 times lower than the \watwo \ts line, which is compatible
with the model (expected ratio of 4).
The water lines are very optically thick, but information comes
from their ratio to the CO lines, which here
is $ I(H_2O)/I(CO) \sim$ 0.5, equal to the ratio observed
in Mrk231 or the Cloverleaf quasar at z=2.6, SDP.17b at z=2.3, and
a little lower than in APM08279 at z=3.91 (van der Werf \etal 2011, and references therein).
In all these galaxies, this ratio is 1-2 orders of magnitude higher than in Orion bar.
 This suggests the existence of a hot dense component, or an XDR in \hls. 

\subsection{The [NII]205\mics line}

Figure \ref{fig:spectra} displays the spectrum of the 
$^3$P$_1$-$>^3$P$_0$ ground state fine structure line of ionized nitrogen, 
smoothed to a resolution of 100\kms.
 The two velocity components are clearly detected, and the blue component
is much broader than for the CO and CI lines.
This fine structure line, with a low excitation potential, mainly traces
HII regions photoionized by OB stars.
In the Milky Way, the line appears moderate, with an intensity ratio
[CII]/[NII] $\sim$ 10. The latter ratio is 20
in M82 (Petuchowski \etal 1994), and 15 in Mrk231 (Fischer \etal 2010).
Only upper limits have been reported
in high-z galaxies, the most constraining being in 
the J1148+5251 quasar at z=6.4 (Walter \etal 2009),
with  L$_{[NII]}$/L$_{FIR}<$ 2 10$^{-5}$, and 
   L$_{[NII]}$/L$_{CO(6-5)}<$ 0.9.
Here we detect the line with 
  L$_{[NII]}$/L$_{FIR}=$ 4 10$^{-5}$ and
  L$_{[NII]}$/L$_{CO(6-5)}=$ 0.15. These values are
intermediate between the AGN-dominated Mrk231, where  L$_{[NII]}$/L$_{FIR}=$ 0.8 10$^{-5}$,
and the starburst M82,  where  L$_{[NII]}$/L$_{FIR}=$ 7.5 10$^{-5}$.
The broad line seen in the [NII] blue velocity component could be related to galactic
winds, often seen in AGN/starburst composite ULIRGs, and with
higher velocities than in neutral flows (Veilleux \etal 2005;
Rupke \etal 2005; Fischer \etal 2010).

%__________________________________________________ One column table
   \begin{table}[!t]
      \caption[]{Line upper limits at 3$\sigma$ in channels of 140\kms}
         \label{tab:limits}
            \begin{tabular}{l c c }
            \hline
            \noalign{\smallskip}
            Line &  $\nu_{obs}$ (GHz)  &  $S_\nu$ (mJy)$^{*}$  \\
  HCN \& HCO$^+$(6--5) & $\sim$ 86.  &  $<4.$ \\
  $^{13}$CO(5--4) & 88.2484   &  $<10.$ \\
   H$_2$O$_o$(1,1,0-1,0,1) &   89.2111  &  $<4.$ \\
  CS(11--10)    & 86.2882  &  $<10.$ \\
%   NII   &  234.0431 &  $<6.$ \\
            \noalign{\smallskip}
            \hline
           \end{tabular}
%\begin{list}{}{}
%\item[$^{*}$] The upper limits are  3$\sigma$ estimates, in channels of 140\kms
%\end{list}
\end{table}
%---------------------------------------------

\subsection{Other lines}

We have searched for the high-density tracers HCN and HCO$^+$,
in their lowest available level, i.e. J=6-5 (Table \ref{tab:limits}). 
The upper limits found
HCN(6-5)/CO(6-5) $<$ 0.25 are not yet constraining, 
since ULIRGs have on average HCN/CO $\sim$ 0.1, except
for 1 or 2 luminous quasars (Gao \& Solomon 2004).

\section{Discussion and conclusion}
\label{discuss}

We have estimated an amplification $\mu \simlt$11 from {\sc lenstool}
(likely range 5-11). The amplification factor can also be estimated from the 
trend of the CO(1-0) $FWHM$ line width with luminosity $L$
(Harris \etal 2012; Bothwell \etal 2012 in prep):
$\mu = 3.5 * (L'/10^{11}K \kms pc^2) * (FWHM/400\kms)^{-1.7}$.
 This yields $\mu$=16$\pm5$, which is consistent with our earlier estimate,
and  suggests that the amplification is likely $\sim$ 11.
Taking into account this magnification, the intrinsic 
FIR luminosity of the source is still in the $\sim$ 10$^{13}$\lsol\,
range, corresponding to a hyper-luminous source. The detected lines
(except water) are divided into two velocity components, which
suggests a galaxy merger, and would explain the
high derived SFR (Engel \etal 2010). 

At these high luminosities, an AGN component
is probable, which is supported by the high water emission.
However, the flux in the CO ladder
peaks at $J=6$ as in starbursts like NGC 253 and M82, and we have no 
evidence yet that it is extended to higher J as in powerful AGNs like Mrk231 or APM08279.
The [NII]205\mics line is detected for the first time at high-z, and its 
relative luminosity with respect to L$_{FIR}$ is intermediate between
a starburst (M82) and an AGN (Mrk231). Its blue velocity component
is much broader than for the CO lines, which could come from
violent flows of ionized gas. The two velocity components have quite different properties
(in H$_2$O, [NII]), favoring an interpretation in terms of 
two galaxies, with a high velocity difference of 540\kms.
Observations of the higher-J CO lines are needed to complete
the gas excitation, and interferometer maps of key lines,
such as the [CII] line, will bring more information about the nature
of this interesting system.

\begin{acknowledgements}
IRAM is supported by INSU/CNRS (France), MPG (Germany) and IGN (Spain).
Herschel is an ESA space observatory with instruments provided
by European-led PI consortia and with important
participation from NASA.
Support for this work was provided in part by NASA through an award issued by JPL/Caltech.
The SMA is a joint project between the Smithsonian
Astrophysical Observatory and the Academia Sinica Institute of Astronomy
and Astrophysics and is funded by the Smithsonian Institution and the
Academia Sinica.
NRAO is operated by Associated Universities Inc., under a cooperative
agreement with the National Science Foundation.
The Keck observatory, made possible thanks to the generous support of
W. M. Keck Foundation, is operated by Caltech, the University of California, and NASA.
\end{acknowledgements}

\end{document}